\documentclass[aps,prl,showpacs,twocolumn]{revtex4}
\usepackage{color}
\usepackage{graphicx}
\usepackage{cancel}

\begin{document}

\title{Nonlinear anomalous Hall effect and negative magnetoresistance \\ in a system with random Rashba field}

\author{V. K. Dugaev$^{1,2,3}$, M. Inglot$^1$, E. Ya. Sherman$^{4,5}$, J. Berakdar,$%
^2$ and J. Barna\'s$^{6,7}$}

\affiliation{$^1$Department of Physics, Rzesz\'ow University of Technology,
al. Powsta\'nc\'ow Warszawy 6, 35-959 Rzesz\'ow, Poland \\
$^2$Institut f\"ur Physik, Martin-Luther-Universit\"at Halle-Wittenberg,
Heinrich-Damerow-Str. 4, 06120 Halle, Germany \\
$^3$Department of Physics and CFIF, Instituto Superior T\'ecnico, TU Lisbon,
Av.~Rovisco Pais, 1049-001 Lisbon, Portugal \\
$^4$Department of Physical Chemistry, Universidad del Pa\'is Vasco UPV-EHU,
48080, Bilbao, Spain \\
$^5$IKERBASQUE Basque Foundation for Science, Alameda Urquijo 36-5, 48011
Bilbao, Spain \\
$^6$Faculty of Physics, Adam Mickiewicz University, Umultowska 85, 61-614
Pozna\'n, Poland \\
$^7$Institute of Molecular Physics, Polish Academy of Sciences,
Smoluchowskiego 17, 60-179 Pozna\'n, Poland}

\begin{abstract}
We predict two spin-dependent transport phenomena in
two-dimensional electron systems, which are induced by spatially
fluctuating Rashba spin-orbit interaction. When the electron gas
is magnetized, the random Rashba interaction leads to the
anomalous Hall effect. An example of such a system is a narrow-gap magnetic
semiconductor-based symmetric quantum well. We show that the
anomalous Hall conductivity reveals a strongly nonlinear
dependence on the magnetization, decreasing exponentially at
large spin density. We also show that electron scattering from a
fluctuating Rashba field in a two-dimensional nonmagnetic electron
system leads to a negative magnetoresistance arising solely due to spin-dependent
effects.
\end{abstract}

\date{\today }
\maketitle

{\it Introduction.} Effects of electron spin on charge transport,
being the basic idea of spintronics, attract a great deal of
interest in fundamental and applied physics as well as in
materials science. There are several origins of the spin
dependence of charge transport; one of them is the spin-orbit (SO)
interaction. This interaction also plays an important role in spin
manipulation in spintronics devices
\cite{zutic04,fabian07,Wu10,dyakonov08}. Of particular importance
is the Rashba SO coupling, usually attributed to quasi
two-dimensional (2D) electron systems in metallic or semiconductor
nanolayers on a substrate,  2D semiconductor heterostructures with
no $z\to -z$ symmetry (with the axis $z$ normal to the layer), or
to surface states \cite{rashba64,pfeffer99,koga02}. This
interaction enables an electrical control of spin precession of 2D
electrons -- the phenomenon utilized in still hypothetical
Datta-Das transistor \cite{datta90}, where current depends on the
angle of spin precession when carriers pass through the device.

Important manifestations of the SO coupling are the anomalous Hall
effect (AHE) and spin Hall effect (SHE)
\cite{jungwirth02,onoda02,sinova04}. In the case of AHE, a charge
current perpendicular to electric field appears without external
magnetic field when the system has a nonzero spontaneous
magnetization \cite{dugaev01}. In the SHE, in turn, static electric field
generates spin current perpendicular to the field orientation.
Both effects, however, can be completely suppressed by
disorder \cite{nunner07,engel07,nagaosa10}.

In symmetrical semiconductor quantum wells, the Rashba SO
interaction vanishes on average. However, spatial fluctuations of
the Rashba coupling may still appear in the system and can play a
qualitatively important role \cite{sherman03,dugaev09,glazov10}.
Surprisingly, in the case of spatially fluctuating Rashba
field, the SHE in 2D electron gas becomes robust to the effect
of impurities, i.e., there is no complete suppression of the
spin-Hall conductivity by disorder \cite{dugaev10}.

In this Letter we predict further two  manifestations of
spin-dependent transport in 2D electron systems with random Rashba
interaction. First, we show that systems with homogeneous
magnetization display AHE. More specifically, we calculate the
off-diagonal conductivity of a 2D magnetized electron gas in a
symmetric quantum well with spatially correlated fluctuations of
SO interaction. We show that the Hall conductivity reveals a very
unusual dependence on the magnetization: it reaches a maximum
followed by a fast decrease to zero for magnetization increasing
further. Second, we calculate the magnetoresistivity of a nonmagnetic
electron gas for a magnetic field parallel to the quantum well and
show that the magnetoresistance is negative. This negative
magnetoresistance (NMR) is a purely classical effect, unrelated to
quantum localization corrections to the conductivity. In both above
mentioned phenomena, the random spin-orbit coupling is crucial and
either generates a macroscopic electric current or modifies the
macroscopic conductivity.

\begin{figure}[h]
\centering
\includegraphics[width=0.30\textwidth]{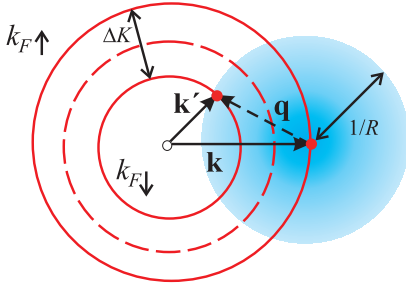}
\caption{Spin-split Fermi surface. $1/R$ is the possible range of
momentum change, $\left|{\bf q}\right|=\left|{\bf k}^{\prime}-{\bf
k}\right|$ in the spin-flip scattering process between two Fermi
surfaces. Dashed circle corresponds to the Fermi momentum $k_{F}$.
In the limit of small Fermi surface splitting, $\Delta K=2Mm/k_{F}$ for $B=0$ (spin orientation is taken with respect to
the $z$-axis), or $\left|\Delta K\right|=2\beta Bm/k_{F}$ for
$M=0$ (spin orientation is taken with respect to the $x$-axis).}
\label{split}
\end{figure}

{\it Model.}
The generic model of a 2D electron gas with random Rashba
interaction is described by the Hamiltonian
$\hat{H}=\hat{H}_{0}+\hat{H}^{\rm (so)}$, where
(we use the units with $\hbar \equiv 1$)
\begin{eqnarray}
&&\hat{H}_{0}=-\frac{\nabla ^{2}}{2m}+U_{\rm rnd}-M\sigma _{z}+\sigma _{x}\beta B,
\label{H0} \\
&&\hat{H}^{\rm (so)}=-\frac{i}{2}\sigma _{x}\left\{ \nabla _{y},\,\lambda (\mathbf{r}%
)\right\} +\frac{i}{2}\sigma _{y}\left\{ \nabla _{x},\,\lambda (\mathbf{r}%
)\right\} .  \label{Hso}
\end{eqnarray}
Here $M$ is half of the spin splitting corresponding to the
homogeneous magnetization along the axis $z$, having purely spin character and not related 
to any conventional macroscopic magnetic field \cite{nagaosa10}. 
The random Rashba coupling parameter $\lambda (\mathbf{r})$   
(for $\mathbf{r}=(x,y))$ has zero average, $\left\langle \lambda
(\mathbf{r})\right\rangle =0$, and a Gaussian correlator
$\langle\lambda(\mathbf{r})\lambda(\mathbf{r^{\prime}})\rangle$,
while $\sigma _{i}$ ($i=x,y,z$) are the spin Pauli matrices and
$m$ is the electron effective mass. The term $U_{\rm rnd}$
describes the spin-independent disorder assumed to be of
white-noise type, $B$ is the in-plane magnetic field along the
axis $x$, and $\beta =g\mu _{B}/2$, where $g$ is the electron
Land\'{e} factor. Assuming in-plane magnetic field we avoid the
diamagnetic orbital effects, and therefore we take into account
only the Zeeman term in Eq.(1). The magnetization leads to
splitting of the Fermi surface, and in the limit of small
magnetization $\Delta K\equiv
k_{F\uparrow}-k_{F\downarrow}=2Mm/k_{F}$, as shown in Fig.~\ref
{split}, where $k_F$ is the Fermi wavevector in the corresponding
nonmagnetic state.

In the momentum representation Eq.~(\ref{Hso}) can be rewritten
as
\begin{equation}
H_{\mathbf{kk^{\prime }}}^{\rm (so)}
=\frac{\lambda _{\mathbf{kk^{\prime }}}}2
\left[
\sigma _{x}(k_{y}+k_{y}^{\prime })-\sigma _{y}(k_{x}+k_{x}^{\prime })
\right],
\label{Hsok}
\end{equation}
where $k_{i}(k_{i}^\prime)$ are the in-plane momentum components ($i=x,y$).
For an external electromagnetic field,
$\mathbf{A}(t)=\mathbf{A}_{0}e^{-i\omega t}$,  one has to make the
following replacement in Eqs.~(\ref{H0}) and (\ref{Hso}): $%
\nabla \to \nabla -ie\mathbf{A}/c$. Thus, the matrix elements of
the coupling to the electromagnetic field  take the form
\begin{eqnarray}
H_{\mathbf{kk^{\prime }}}^{(A)}
&=&\left[ -\frac{e}{mc}\,\left( \mathbf{k}%
\cdot \mathbf{A}\right) +\frac{e^{2}}{2mc^{2}}\,A^{2}\right] \delta _{%
\mathbf{kk^{\prime}}} \nonumber \\
&&-\frac{e}{c}\lambda_{\mathbf{kk^\prime }}
\left(\sigma _{x}A_{y}-\sigma _{y}A_{x}\right),
\label{6}
\end{eqnarray}
and include the term following from the random Rashba
interaction. Correspondingly, the matrix elements of the charge current operator
$\hat{\mathbf{j}}=-c\,\partial \hat{H}^{(A)}/\partial \mathbf{A}$
are
\begin{equation}
(j_{x,y})_{\mathbf{kk^{\prime }}}=
\frac{e}{m}\,\left( k_{x,y}-\frac{e}{c}A_{x,y}\right)\delta_{\mathbf{kk^{\prime }}}
\mp e\lambda_{\mathbf{kk^{\prime }}}\sigma _{y,x}.
\label{jxy}
\end{equation}
The random spin-orbit coupling leads to spin-flip scattering
between states with opposite spin orientations. This, in turn,
results in two effects to be analyzed below: the AHE and the NMR.

\begin{figure}[h]
\includegraphics[width=0.30\textwidth]{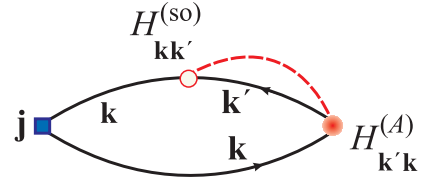}
\caption{Feynman graph for the anomalous Hall current. Matrix elements in the vertices
are determined by Eqs.(\ref{Hsok})-(\ref{jxy}).}
\label{diagrams}
\end{figure}

{\it Anomalous Hall effect.} Let us begin with the AHE induced by fluctuating Rashba field. Hence, we put here
$B=0$, assume an electric field $\mathbf{E}$ along the axis $y$,
and consider charge current normal to the field. Upon calculating
the off-diagonal linear conductivity \cite{agd}, we find that it
can be written as the difference of the contributions from two
different spin channels,
\begin{equation}
\sigma _{xy}=\sigma _{xy\uparrow }-\sigma _{xy\downarrow }\,.
\label{difference}
\end{equation}
In the approximation corresponding to the loop diagrams shown in
Fig. \ref {diagrams} one finds ($s\ne s^{\prime }$)
\begin{eqnarray}
\sigma _{xy,\,s} &=&\frac{e^{2}}{m}\sum_{\mathbf{kq}}\,|\lambda
_{q}|^{2}\,k_{y}(2k_{y}-q_{y})\, \delta (\mu -\varepsilon _{\mathbf{k-q},s^{\prime }})
\nonumber  \label{10} \\
&&\times G_{k,s}^{R}\,G_{k,s}^{A}\,,  \label{sigma:xy:s}
\end{eqnarray}
where $G_{k,s}^{R}$ and $G_{k,s}^{A}$ ($s=\uparrow ,\downarrow$)
are the retarded and advanced Green functions, respectively.  Due
to scattering from the fluctuating SO field, electrons with
opposite spins are turned to the opposite transverse directions,
so  $\sigma _{xy\uparrow }$ and $\sigma_{xy\downarrow}$ enter
Eq.(\ref{difference}) with opposite signs (they are added in the
case of spin Hall effect). The resulting AHE for $M\ne 0$ is
nonzero due to the spin polarization (magnetization) of the
electron gas.

\begin{figure}[h]
\includegraphics*[width=0.45\textwidth]{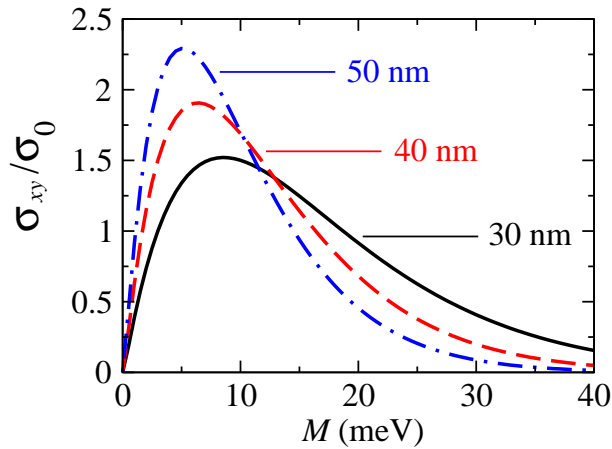}
\caption{Anomalous Hall conductivity as a function of spin
splitting parameter $M$ for indicated values of $R$. Other
parameters are defined in the text.} \label{ahe}
\end{figure}

Upon calculating the sum  over $k$ in Eq.~(\ref{sigma:xy:s})
one obtains
\begin{eqnarray}
\sigma _{xy\uparrow ,\downarrow } &=&\frac{2e^{2}m\nu \tau _{\uparrow
,\downarrow }}{\pi k_{F\uparrow ,\downarrow }}\int\limits_{0}^{\infty
}dq\left( \mu -\frac{q^{2}}{8m}\pm \frac{3M}{2}\right) |\lambda _{q}|^{2}
\nonumber  \label{11} \\
&&\times \int\limits_{0}^{\pi }d\varphi \,\delta \left( \cos \varphi -\frac{q%
}{2k_{F\uparrow ,\downarrow }}\mp \frac{2mM}{k_{F\uparrow ,\downarrow }q}%
\right) ,
\end{eqnarray}
where $\tau _{\uparrow ,\downarrow }$ is the spin-dependent
relaxation time, $\nu $ is the density of states per spin, and the
upper (lower) sign corresponds to spin-up (spin-down) electrons.
The spin-dependence of the relaxation time $\tau _{\uparrow
,\downarrow }$ can result from the spin dependence of the
corresponding Fermi momentum. In the case of white-noise short
range disorder, $\tau $ is determined solely by the density of
states at the Fermi level, and, therefore, does not depend on the
spin orientation.

If the magnetization is relatively weak, $\sqrt{2mM}<k_{F}$, both spin
subbands are occupied and the spin-projected conductivity is
\begin{eqnarray}
&&\sigma _{xy\uparrow ,\downarrow }=\frac{2e^{2}m\nu \tau _{\uparrow
,\downarrow }}{\pi k_{F\uparrow ,\downarrow }}  \\
&&\times \int\limits_{\Delta K}^{k_{F\uparrow}+k_{F\downarrow}}dq|\lambda _{q}|^{2}
\frac{\mu -q^{2}/8m\pm 3M/2}{\sqrt{%
1-(q/2k_{F\uparrow ,\downarrow }\pm 2mM/k_{F\uparrow ,\downarrow }q)^{2}}}.\nonumber
\label{sigma:result}
\end{eqnarray}
In turn, when the magnetization is sufficiently strong,
$\sqrt{2mM}>k_{F}$, only one spin subband is occupied.
Spin-flip scattering is then absent and $\sigma _{xy}$ vanishes.
As we show below in more details, dependence of the AHE on the
magnetization $M$ is rather unusual.

For further calculations we need a more specific Rashba field
correlator, and we  assumed it in the following generic form:
\cite {glazov10}
\begin{equation}
|\lambda _{q}|^{2}=2\pi \langle \lambda ^{2}\rangle R^{2}e^{-qR}.
\label{correlator}
\end{equation}
{Here $\left\langle \lambda ^{2}\right\rangle $ characterizes
the field variation, and the correlation length $R$ is
of the order of the distance between the quantum well and the dopant layers since 
the in-plane distribution of electric field in the well is controlled by this parameter \cite{ando82}.}

Using Eqs.~(\ref{sigma:result}) and (\ref{difference}), we have calculated
numerically the Hall conductivity as shown in
Fig.~\ref{ahe}. The conductivity is presented there in the units
of $\sigma _{0}=\sigma _{D}\left\langle \lambda ^{2}\right\rangle
k_{F}^{2}/E_{F}^{2}$, where  $\sigma _{D}=ne^{2}\tau /m$ is the
Drude conductivity, and $E_{F}=k_{F}^{2}/2m$. We used the
parameters characteristic for InSb: effective mass
$m=0.0134\,m_{0}$, and electron density $n=5\times
10^{11}$~cm$^{-2}$ related to the Fermi momentum $k_{F}=(2\pi
n)^{1/2}$.  We can estimate the ratio of $\sigma _{0}/\sigma _{D}$ by using the relation
between the electric field $\mathcal{E}$ and the spin-orbit coupling in the form $%
\left\langle \lambda ^{2}\right\rangle =\xi ^{2}e^{2}\left\langle \mathcal{E}%
_{r}^{2}\right\rangle ,$ where the variation in the random electric field $%
\left\langle \mathcal{E}_{r}^{2}\right\rangle =2\pi
e^{2}n_{d}/\kappa ^{2}R^{2}.$ For InSb $\xi\approx 5\mbox{nm}^2$,\cite{Winkler,Morgenstern} 
and for typical parameters of quantum wells we obtain 
$\sqrt{\langle\lambda^{2}\rangle}$ of the order of $10^{-6}$ meVcm and
$\sigma _{0}/\sigma _{D}$ of the order of
10$^{-2}-10^{-1}$. Thus, the maximum anomalous Hall conductivity
can be of an order of 0.01-0.1 of the Drude conductivity.

As follows from Fig.~\ref{ahe}, the Hall conductivity has a sharp
maximum at a certain value of the parameter $M$. The physical
reason for such a behavior is that the spin-flip scattering from
fluctuating Rashba coupling is effective only for a relatively
small change in electron momentum, $q<1/R$. Thus, if $\Delta
K>1/R$, these elastic spin-flip processes become suppressed and
the AHE vanishes. To have a better physical insight into the
problem, we present the Hall conductivity in an approximate form
as
\begin{equation}
\frac{\sigma _{xy}}{\sigma _{0}}\sim k_{F}^{2}R^{2}\frac{M}{E_{F}}\exp \left(
-\frac{M}{E_{F}}k_{F}R\right).  \label{approximation}
\end{equation}
At zero $M$ the contributions from different spins exactly
compensate each other, while with the increase in $M$ to the
region where $\Delta{K}R=\left({M}/{E_{F}}\right)k_{F}R\gg1$, the spin-flip transitions are
suppressed and the Hall conductivity vanishes as well. The maximum
of conductivity is achieved for  ${M}={E_{F}}/k_{F}R$. The
resulting maximum anomalous Hall conductivity is then of the order
\begin{equation}
\max \left(\frac{\sigma _{xy}}{\sigma _{0}}\right) \sim k_{F}R/e,
\label{maximum}
\end{equation}
where $e\approx2.718\dots$ is the base if the natural logarithm.

It is instructive to compare the anomalous and the conventional Hall
conductivities. Magnetization of the order of ${E_{F}}/k_{F}R$ for
InSb with the $g-$factor close to $50$ can be achieved in the
fields $B_{z}$ of the order of 1 T. The ratio of the Drude
and Hall conductivity is $\omega _{c}\tau,$ where $\omega _{c}$
is the cyclotron frequency. In the field of 1 T for a
sample with the mobility of 10$^{5}$ cm$^{2}$/Vs, one finds
$\omega_{c}\tau \approx 10.$ As a result, the anomalous Hall
conductivity can be of an order of $0.1-1.0$ of the conventional
Hall conductivity. It can be extracted from the conventional Hall
measurements taking into account an unusual dependence on the
external magnetic field.

{\it Negative magnetoresistance and anisotropic conductivity.}
We have shown above that the spin-flip scattering from fluctuating
spin-orbit field leads to AHE in a magnetized electron gas.
Now we demonstrate that such scattering also
modifies the diagonal  conductivity as the electron relaxation
rate includes generally not only the term due to scattering from
impurities but also the contribution from spin-flip scattering due
to fluctuating Rashba coupling. Since the spin-flip scattering is
essential at small spin splitting and decreases at large
splitting, its contribution to spin relaxation time strongly
depends on the spin polarization of the electron gas. This spin
polarization can be modified by an external magnetic field, which
in turn leads to the magnetoresistance. For this effect it is
sufficient to consider the situation when the band splitting is
not related to the homogeneous magnetization but is due to the
Zeeman field. Therefore we consider now a nonmagnetic electron
gas, $M=0$,  and assume a nonzero magnetic field along the $x$
axis. We show below that spin-orbit fluctuating Rashba field leads
then to a NMR.

To calculate the diagonal component of conductivity tensor we can
use the kinetic equation, which takes into account spin-conserving
and spin-flip transitions,
\begin{eqnarray}
\label{kinetic}
\frac{e}{m}
\left(\mathbf{E}\cdot\mathbf{k}\right)
\frac{\partial f_{\mathbf{k}\uparrow
}^{0}}{\partial \varepsilon_{k}} &=&-\sum_{\mathbf{k^{\prime }}}
\Large[W_{\mathbf{kk^{\prime }}}(f_{\mathbf{k}\uparrow }-f_{\mathbf{k^{\prime }}%
\uparrow })\nonumber  \label{16} \\
&&+W_{\mathbf{kk^{\prime }}}^{f}(f_{\mathbf{k}\uparrow }-f_{\mathbf{%
k^{\prime }}\downarrow })\Large],
\end{eqnarray}
and a similar equation for the opposite spin orientation. Here we
introduced the notation
\begin{eqnarray}
\label{Wconserv}
&&W_{\mathbf{kk^{\prime }}}=2\pi \left[ w_{\mathbf{kk^{\prime }}}+|\lambda _{%
\mathbf{kk^{\prime }}}|^{2}(k_{y}+k_{y}^{\prime })^{2}\right] \delta
(\varepsilon _{k}-\varepsilon _{k^{\prime }}),\hskip0.5cm
\\
\label{Wflip}
&&W_{\mathbf{kk^{\prime }}}^{f}=2\pi |\lambda _{\mathbf{kk^{\prime }}%
}|^{2}(k_{x}+k_{x}^{\prime })^{2}\delta (\varepsilon
_{k}-\varepsilon _{k^{\prime }}\pm 2\beta B),
\end{eqnarray}
for spin-conserving and spin-flip transitions, respectively, where $w_{%
\mathbf{kk^{\prime }}}$ corresponds to the spin-conserving
scattering from usual disorder. If the potential of impurities is
short range, $w_{\mathbf{kk^{\prime }}}$ can be taken independent
on the momentum, $w_{\mathbf{kk^{\prime }}}\simeq w_{0}$. Although the solution
of Eq.(\ref{kinetic}) can in general be presented as a spin-dependent sum of cylindrical Fermi
surface harmonics, it cannot be solved in the general form due to anisotropy of matrix elements
in Eqs.(\ref{Wconserv}),(\ref{Wflip}) and the presence of two Fermi surfaces.
However, for $k_{F}R\gg1$, where one can with a high accuracy
put $k_{x}=k_{x}^{\prime}$ and $k_{y}=k_{y}^{\prime}$,
they can be greatly simplified for the
electric field $\mathbf{E}$ parallel and perpendicular to the magnetic field $\mathbf{B}$.

Let us  start with the limit of weak magnetic field, when the
Zeeman splitting is very small, $\Delta K\ll 1/R.$ The spin-split
Fermi surfaces are then almost identical, and can be considered as
a single surface. The corresponding isotropic conductivity,
$\sigma =\sigma_D+\delta\sigma^{\rm so}$, can be also presented as
$\sigma (B\to 0)=ne^2\tau_{\rm tot}/m$, where $1/\tau_{\rm
tot}=1/\tau+1/\tau_{\rm so}$ includes the transport scattering
rate $1/\tau_{\rm so}$ related to Eqs. (\ref{Wflip}) (spin-flip
processes) and (\ref{Wconserv}) (spin-conserving processes caused
by spin-orbit coupling). As a result, $\delta\sigma^{\rm
so}=-\tau\sigma_{D}/\tau_{\rm so}$. The rate $1/\tau_{\rm so}$ can
be evaluated as $1/\tau_{\rm so}\sim 4\left\langle \lambda
^{2}\right\rangle /v_{F}R$ and is of the order of 10$^{11}$
s$^{-1}$, leading to the negative correction of the order of
0.1$\sigma _{D}$ to the Drude conductivity at mobility 10$^{5}$
cm$^{2}$/Vs.

When the magnetic field increases, the spin-flip processes become suppressed,
as can be qualitatively seen in Fig.\ref{split}, while spin-conserving ones remain
almost intact. In the limit of strong field, $\beta BmR\gg 1$, the spin-flip term,
Eq.~(\ref{Wflip}), vanishes and the conductivity is
larger than at $B=0$ for both ${\mathbf E}\parallel{\mathbf B}$ and ${\mathbf E}\perp{\mathbf B}$
geometries. At the same time, as we see from
(\ref{Wconserv}), the remaining spin-conserving processes include scattering
from the random Rashba field that strongly depends on the
orientation of momenta ${\bf k}$ and ${\bf k'}$. As a result, the
total scattering at large field becomes anisotropic. Solving the
kinetic equation for this case we find that the conductivity at
large $B$ is also anisotropic with
$\left|\sigma\left({\mathbf E}\parallel{\mathbf B}\right)-\sigma\left({\mathbf E}\perp{\mathbf B}\right)\right|=\left|\delta\sigma^{\rm so}/2\right|$,
and the degree of anisotropy is of the order of several percent.

Note that the NMR effect we consider here is qualitatively
different from the one of Ref.~\cite{Mirlin} for systems with
long- and short-range disorder. The mechanism proposed by us has a
solely spin-related origin and is not related to the orbital
motion considered in Ref.~\cite{Mirlin}. Recent experiments
\cite{Zudov} showed that the NMR cannot be fully explained in
terms of the model of Ref.[\onlinecite{Mirlin}] and some
spin-related effects can be involved in the physics of this
phenomenon. Moreover, our effect is purely classical and not
related to NMR in the weak localization regime.\cite{Faniel,Kohda}

{\it Summary.} We have calculated the anomalous Hall conductivity
and the magnetoresistance of 2D electron systems due to scattering
from spatial fluctuations of Rashba spin-orbit interaction. The
materials where this interaction can be essential are symmetric
narrow-gap semiconductor quantum wells. One of the usual
characterization methods for these materials is based on the
measurements of AHE under the assumption that AHE is proportional
to magnetization. This proportionality, however,  is completely
destroyed when the AHE is related to the fluctuating Rashba field.
It should be stressed that the proposed mechanism of AHE can be
important when the usual impurity scattering within the quantum
well is small. In the case of nonmagnetic semiconductors it can be
easily realized by donor impurities outside the quantum well. We
have also demonstrated that the scattering of electrons from the
fluctuations of Rashba field in nonmagnetic symmetric
semiconductor quantum wells leads to a negative magnetoresistance.
This effect can be realized for example in InSb or other narrow
gap semiconductors, in which the Zeeman splitting and spin-orbit
coupling are both very strong.

This work is partly supported by the DFG in Germany
and by National Science Center in Poland as a research project in years 2011-2014 and 
the project DEC-2012/04/A/ST3/00372. The work of EYS was supported by the MCINN of Spain grant FIS
2009-12773-C02-01, "Grupos Consolidados UPV/EHU del Gobierno Vasco" grant
IT-472-10, and by the UPV/EHU under program UFI 11/55.

\end{document}